\documentclass[conference]{IEEEtran}
\IEEEoverridecommandlockouts
\usepackage{cite}
\usepackage{amsmath,amssymb,amsfonts}
\usepackage{algorithmic}
\usepackage{graphicx}
\usepackage{textcomp}
\usepackage{xcolor}
\usepackage{bm}

\def\BibTeX{{\rm B\kern-.05em{\sc i\kern-.025em b}\kern-.08em
    T\kern-.1667em\lower.7ex\hbox{E}\kern-.125emX}}
\begin{document}

\title{Simulating reaction time for Eureka effect in visual object recognition using artificial neural network\\
\thanks{
We thank Drs. Toshio Yanagida, Kunihiko Kaneko, Izumi Ohzawa, Toshiyuki Kanoh, Hideki Kashioka, Tetsuya Shimokawa, and Mr. Keigo Nishida for helpful advice and discussions. This work was supported in part by MIC under a grant entitled ``R\&D of ICT Priority Technology (JPMI00316)''.
}
}

\author{

\IEEEauthorblockN{Kazufumi Hosoda*}
\IEEEauthorblockA{
\textit{Center for Information and} \\
\textit{Neural Networks (CiNet), Advanced}\\
\textit{ICT Research Institute, NICT}\\
Suita, Osaka, Japan \\
hosodak@nict.go.jp}

\and

\IEEEauthorblockN{Shigeto Seno}
\IEEEauthorblockA{
\textit{Department of Bioinformatic } \\
\textit{Engineering, Graduate School }\\
\textit{of IST, Osaka University}\\
Suita, Osaka, Japan \\
senoo@ist.osaka-u.ac.jp}

\and

\IEEEauthorblockN{Tsutomu Murata}
\IEEEauthorblockA{
\textit{Center for Information and} \\
\textit{Neural Networks (CiNet), Advanced}\\
\textit{ICT Research Institute, NICT}\\
Suita, Osaka, Japan \\
benmura@nict.go.jp}
}

\maketitle

\begin{abstract}
The human brain can recognize objects hidden in even severely degraded images after observing them for a while, which is known as a type of Eureka effect, possibly associated with human creativity. A previous psychological study suggests that the basis of this "Eureka recognition" is neural processes of coincidence of multiple stochastic activities. Here we constructed an artificial-neural-network-based model that simulated the characteristics of the human Eureka recognition.

\end{abstract}


\section{Introduction}

Understanding and reconstructing human creativity in terms of neural modeling is important for constructing creativity support systems\cite{Wang}. Eureka (or Aha!) effect is thought to be associated with the human creativity. As a phenomenon in a visual Eureka effect, humans can recognize objects hidden in even severely degraded images (such as Mooney images) suddenly after observing them for a while. Once recognized, the objects remain clear in the conscious awareness. 

In a previous psychological study of this ``Eureka Recognition (ER)'' \cite{Murata}, reaction time (RT) of human subjects to recognize hidden objects in degraded binary (black-and-white) images followed a certain equation in which the RT is determined by both ``individual ability'' and ``image difficulty that varies across images.'' This quantitative relation was explained by assuming that the human brain stochastically activated missing components of a hidden object in a degraded image and ER occurred when multiple missing components were coincidentally activated. A coarse-grained mathematical model that repeats stochastic activation until the coincidence occur, using psychologically-measured image difficulty, explained the following properties of the psychological results: (i) RTs over subjects for a degraded image followed a normal distribution on a logarithmic time scale, (ii) means and standard deviations of the above lognormal distributions over various degraded images showed a linear relation, and (iii) image difficulty showed a discrete distribution characterized by natural numbers. However, this coarse-grained model does not include specific neural mechanisms of ER and how ER is implemented by artificial neural network (ANN) models.

In this study, we constructed an ANN-based model to simulate the ER properties observed in the previous human study\cite{Murata}. Being applied to the same degraded images used in the human experiment, our model successfully reproduced the above three characteristics of ER. Our model can show differences and similarities between the recognition of human and a certain ANN, and it will contribute to the elucidation of the mechanism of Eureka effect and human creativity.

\section{Methods}

\subsection{Dataset and ANN image classifier}

For input images, we used 90 binary images and their corresponding 90 original color images which were also used in the previous human experiment\cite{Murata}. For an ANN image classifier, we used ViT-B/32 (a Vision Transformer\cite{Dosovitskiy}) pretrained with ImageNet 1000 classes. Some of the 90 images were not included in the 1000 classes, and we added 15 new classes to the ANN by DONE (Direct ONE-shot learning\cite{Hosoda}), using three Creative-Commons-license images on the internet as the training data for each of the 15 classes. We determined ``correct class'' for each binary image as a coarse-categorically-correct class that were top-1 or top-2 output obtained when inputting the corresponding color image.

\subsection{ANN-based Model for Eureka Recognition}

To construct an ANN-based model that simulates RT for ER, we integrated the following three parts: an ANN image classifier, a recognition criterion based on multiple neural factors, and a stochastic process. Note that our model was not for classifying a binary image but for obtaining RT, and the model considered only the correct class because in the human experiment the wrong answer was ignored. 
The ANN classifier has a final dense layer, in which the input is a feature vector $\bm{x}$ with $M$ neurons and the output is a vector $\bm{y}$ for $N$ classes that represents the fitness rating to each class of the output by means of a weight matrix $\bm{W}$ ($N \times M$). The rating for $i$-th class $y_i$ ($i=1,2,\cdots,N$) is proportional to the cosine similarity between the feature vector $\bm{x}$ (obtained from image) and the corresponding weight vector $\bm{w}_i$ ($i$-th row vector of $\bm{W}$). Therefore, the weight vector for the correct class $\bm{w}_c$ indicates neurons to fire or rest for the class.

For the recognition criterion, we defined ``specific-neurons-to-fire'' and ``specific-neurons-to-rest'' of the feature vector $\bm{x}$ as neurons whose corresponding weight values were respectively top 5\% and bottom 5\% in the weight vector $\bm{w}_c$. Then, we obtained the sum of the numbers of firing specific-neurons-to-fire (top 5\%) and resting specific-neurons-to-rest (bottom 5\%) in the feature vector $\bm{x}$, which we denote as $K$. Setting a threshold number $K_{\rm{thr}}$, we regarded the input image as being recognized when $K \geq K_{\rm{thr}}$. The threshold value was obtained in common for all 90 binary images as the minimum value of $K$ among 90 color images ($K_{\rm{thr}}=10$), because human subjects recognized all color images instantly and ER did not occur in the human experiment.

When $K$ was less than $K_{\rm{thr}}$, the input image was not yet recognized, and the following stochastic process was introduced. Each of $M$ neurons in $\bm{x}$ randomly fired, rested, or retained the value obtained from the input image, with a probability of $p$, $p$, or $1-2p$, respectively. At each time step, $K$ was evaluated, and this stochastic process was repeated while $K$ was less than $K_{\rm{thr}}$. When $K$ satisfied the threshold, i.e., one or multiple missing components were coincidentally activated, the stochastic process was terminated, and the RT for ER was obtained as the number of time steps of this stochastic process.

\section{RESULTS AND DISCUSSION}

The ANN-based model was applied to obtain RT for the 90 binary images, which were used in the previous human study\cite{Murata}. To validate the model in comparison with the human results, we investigated if the model reproduced the properties of human data described as (i) to (iii) in Introduction section.

\begin{figure}[htbp]
\centerline{\includegraphics[width=8.5cm]{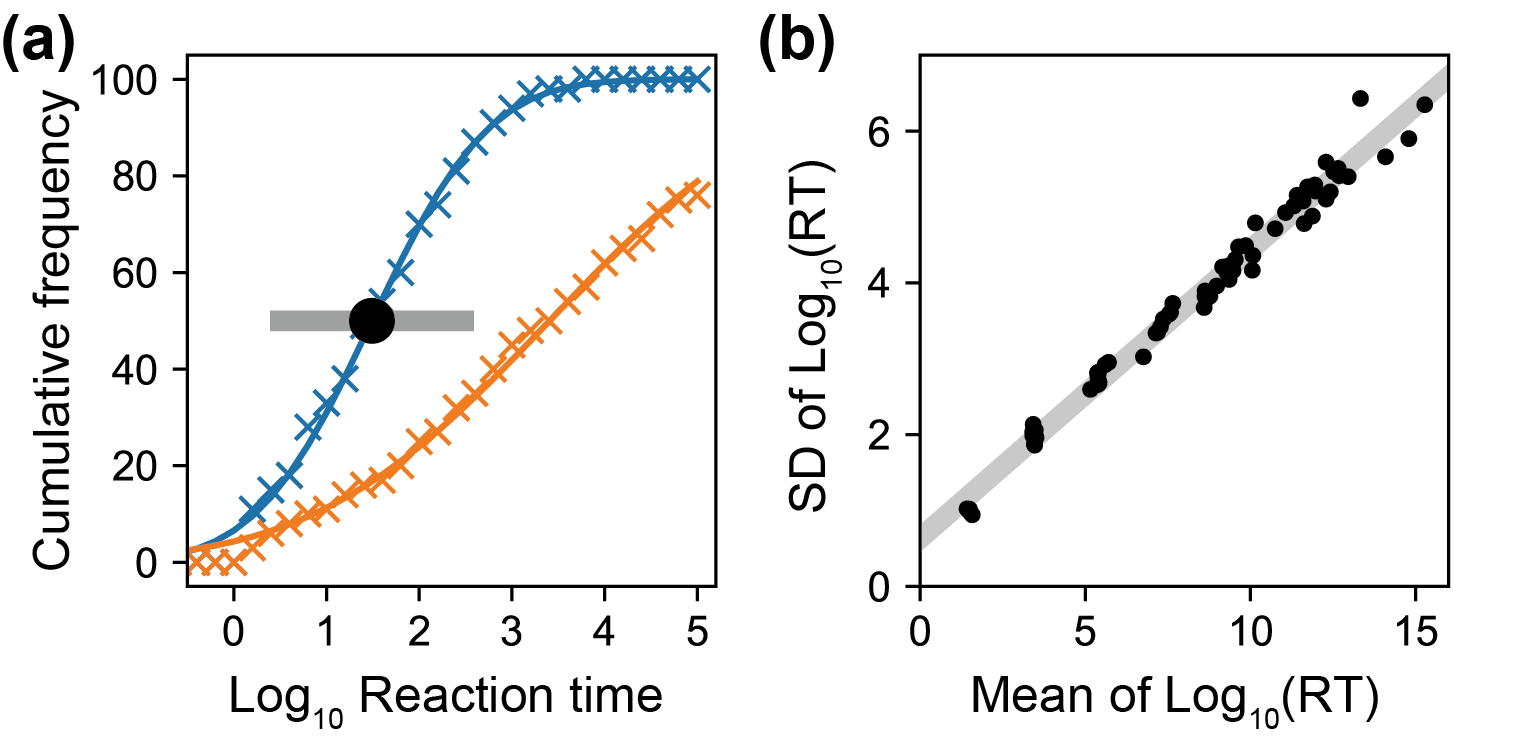}}
\caption{Statistical characteristics. See text for explanation.}
\label{fig:MvsSD}
\end{figure}

We made the firing probability $p$ in the stochastic process (see Methods) to distribute lognormally (according to the human results), and RTs were obtained by simulations using 100 different $p$ values (corresponding to 100 human subjects). Figure~\ref{fig:MvsSD}(a) shows cumulative distributions of the common logarithm of RT with the 100 $p$ values for inputs of two example binary images (blue and orange). These distributions (crosses) were well fitted by normal distributions (lines) as the property (i) of human study (see Introduction), providing values of the logarithmic means and SDs (as indicated by the black circle and gray bar for blue example).

Figure~\ref{fig:MvsSD}(b) shows the relationships between the means and SDs obtained by the model using the 90 binary images. The result clearly shows a linear relationship between the means and SDs over the 90 images ($R=0.99$), successfully reproducing the property (ii) of human study.

\begin{figure}[htbp]
\centerline{\includegraphics[width=8.5cm]{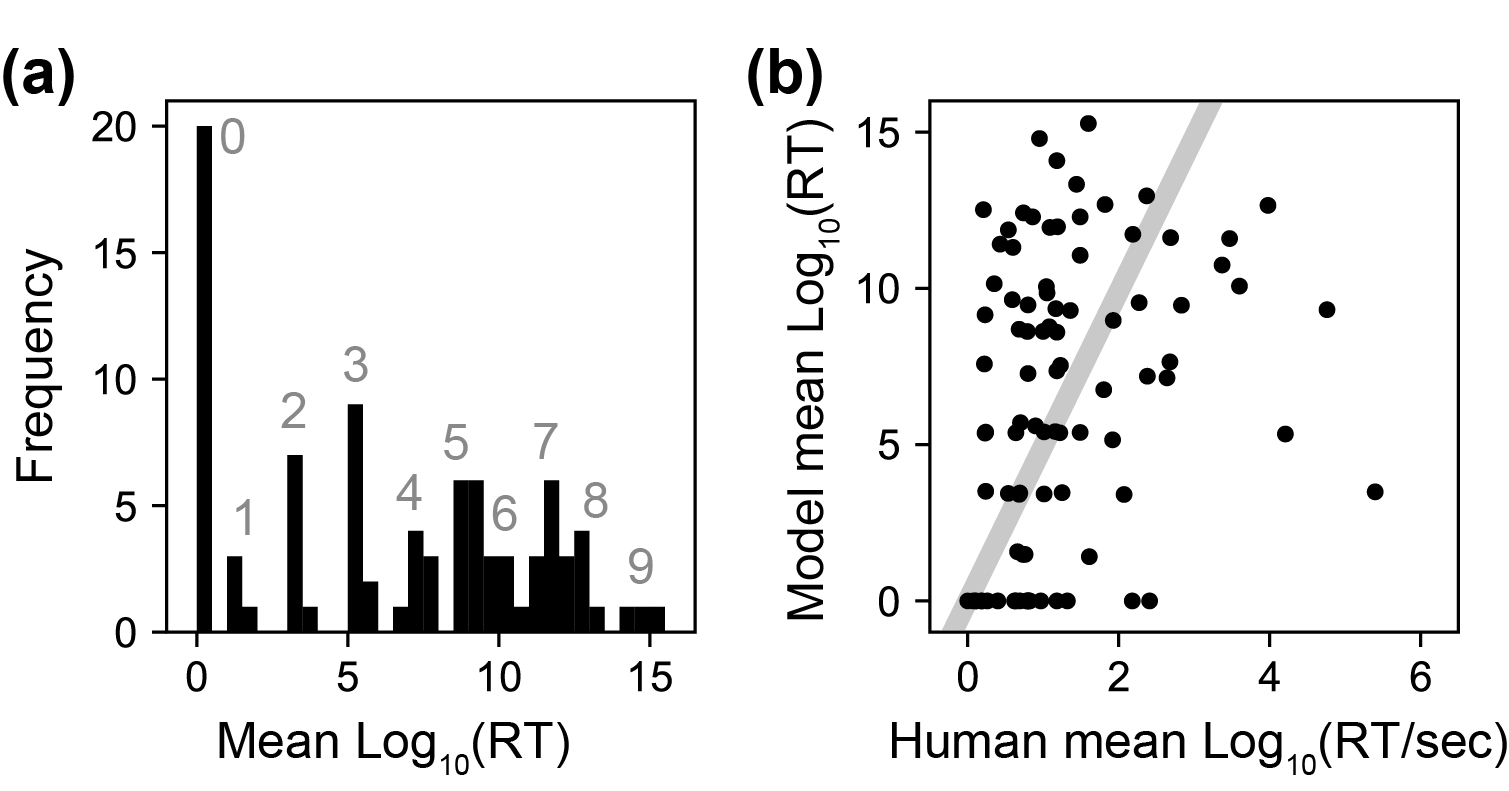}}
\caption{Difference between images. See text for explanation.}
\label{fig:MANvsAI}
\end{figure}

Figure~\ref{fig:MANvsAI}(a) shows a frequency distribution of the mean log of RT of the 90 binary images in our model. The mean log of RT, which can be interpreted as difficulty of recognizing the image, distribute in a discrete manner (as bars with integers). This result successfully provided the corresponding property (iii) of human study.

Figure~\ref{fig:MANvsAI}(b) quantitatively shows the relationship between the human experiment and our model simulation about the mean log of RTs of 90 binary images. There are large differences with a small significant correlation ($R=0.28, P=0.007; \alpha=0.05$) between them, possibly illustrating differences in ways of coding visual objects to neural activities, which is still under elucidation across research fields.

\section{CONCLUSION}

We simulated the reaction time for the Eureka effect in visual object recognition of degraded difficult-to-recognize binary images by using ANN-based model. Our model well reproduced the characteristics of the previous human experimental results\cite{Murata} and quantitatively showed the difference between human and the ANN-based model in image difficulty. The coincidence of multiple stochastic neural activities, which plays an essential role in our ANN-basd model, seems to be consistent with a conventional theory that creativity arises from the association of multiple seemingly independent processes\cite{Wang}. Our model may contribute to the elucidation of the mechanism of creativity and construction of creativity support systems.






\end{document}